\documentclass[journal=jctcce,manuscript=article]{achemso}
\usepackage[utf8]{inputenc}
\usepackage[english]{babel}
\usepackage{amsmath}
\usepackage{amsfonts}
\usepackage{amssymb}
\usepackage{makeidx}
\usepackage{float}
\usepackage{caption}
\usepackage{subcaption}
\usepackage{afterpage}
\usepackage{mathtools}
\usepackage{multirow}
\usepackage[usenames,dvipsnames,svgnames,table]{xcolor}
\newcommand{\onlinecite}[1]{{\citenum{#1}}}

\author{M. Rodr\'iguez-Mayorga}
\email{marm3.14@gmail.com}
\affiliation{Department of Theoretical Chemistry, Vrije Universiteit Amsterdam, De Boelelaan 1083, 1081 HV Amsterdam, The Netherlands.}
\author{D. Keizer}
\affiliation{Department of Theoretical Chemistry, Vrije Universiteit Amsterdam, De Boelelaan 1083, 1081 HV Amsterdam, The Netherlands.}
\author{K.J.H. Giesbertz}
\affiliation{Department of Theoretical Chemistry, Vrije Universiteit Amsterdam, De Boelelaan 1083, 1081 HV Amsterdam, The Netherlands.}
\author{L. Visscher}
\affiliation{Department of Theoretical Chemistry, Vrije Universiteit Amsterdam, De Boelelaan 1083, 1081 HV Amsterdam, The Netherlands.}

\date{\today}
\title{Relativistic effects on electronic pair densities: a perspective from the radial intracule and extracule probability densities.}

\setkeys{acs}{articletitle = true}

\begin{document}
While the effect of relativity in the electronic density has been widely studied, the effect on the pair probability, intracule, and extracule densities has not been studied before. Thus, in this work, we unveil new insights related to changes on the electronic structure caused by relativistic effects. Our numerical results suggest that the mean inter-electronic distance is reduced (mostly) due to scalar-relativistic effects. As a consequence, an increase of the electron-electron repulsion energy is observed. Preliminary results suggest that this observation is also valid when electronic correlation effects are considered.

\newpage

\section{Introduction}
The advent of quantum mechanics lead to a new manner to study systems by describing them in terms of wavefunctions~\cite{schrodinger1926undulatory,schrodinger1926quantisierung}. Nowadays, the wavefunction ($\Psi$) that describes the electronic structure is considered as the fundamental quantity to study (several) physical and chemical processes~\cite{szabo2012modern,helgaker:00book,levine2000quantum}. Nevertheless, the full knowledge of $\Psi$ is not required to compute most quantities of practical interest (e.g.\ dipole moments, bond orders, among others). For these purposes, so-called reduced quantities (RQs), that can be defined from $\Psi$, contain sufficient information. Since the determination of (an accurate) $\Psi$ is computationally very demanding, methodologies based on direct search of RQs are evolving fast in the scientific community, both in non-relativistic~\cite{hohenberg:64pr,kohn:64pr,kohn:65pr,gilbert:75prb,dumont2007intracule,crittenden2007intracule,crittenden2011intracule,gill2011intracule,hollett2011intracule} and in relativistic~\cite{chaix1989quantum,chaix1989quantum2,toulouse2021relativistic,rodriguez2022relativistic} incarnations.   

Already the non-relativistic $\Psi$ is a complicated object that involves $3N$ spatial coordinates for a $N$-electron system in the usual 3D space. This high dimensionality implies, in general, a non-trivial analysis (and visualization) of the information contained in $\Psi$. In contrast to the wavefunctions themselves, the lower dimensional RQs, like the electronic density 
\begin{align}
n({\bf r})= N \int d{\sigma}d{\bf x}_2\dotsi d{\bf x}_N\Psi ^* ({\bf x},{\bf x}_2,\dotsc,{\bf x}_N)  \Psi({\bf x},{\bf x}_2,\dotsc,{\bf x}_N) ,
\label{Eq:edensity}
\end{align}
with ${\bf x}=({\bf r}\sigma)$ and $\sigma =( \uparrow,\downarrow)$, can directly be compared between different electronic structure methods. Additionally, in many cases RQs like the probability density can also be measured experimentally (see for example Refs.~\cite{debye1915scattering,coopens:97book,watanabe:watanabemp2004}), which enhances their applicability and the interest in using them. 

\begin{figure}[H]
\centering
\begin{subfigure}[t]{\textwidth}
\centering
\includegraphics[scale=0.8]{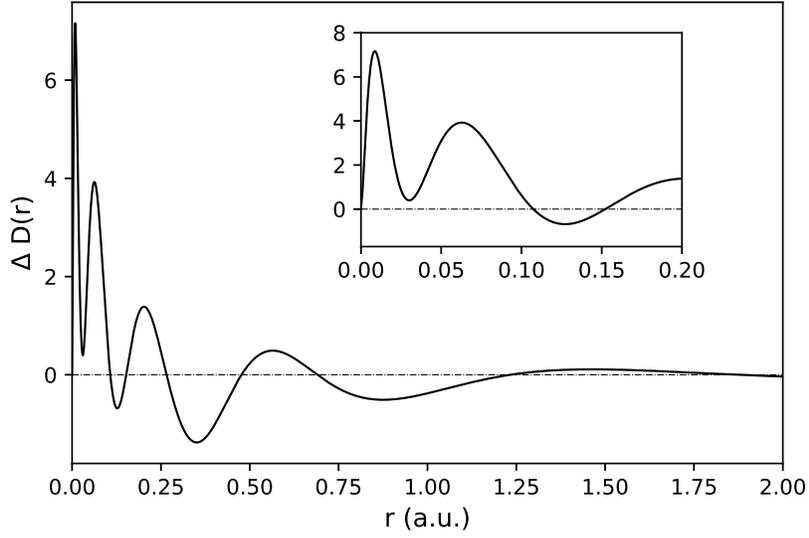}
\caption{}
\label{fig:xe_cg_radialdensitya}
\end{subfigure}
\hfill
\begin{subfigure}[t]{\textwidth}
\centering
\includegraphics[scale=0.8]{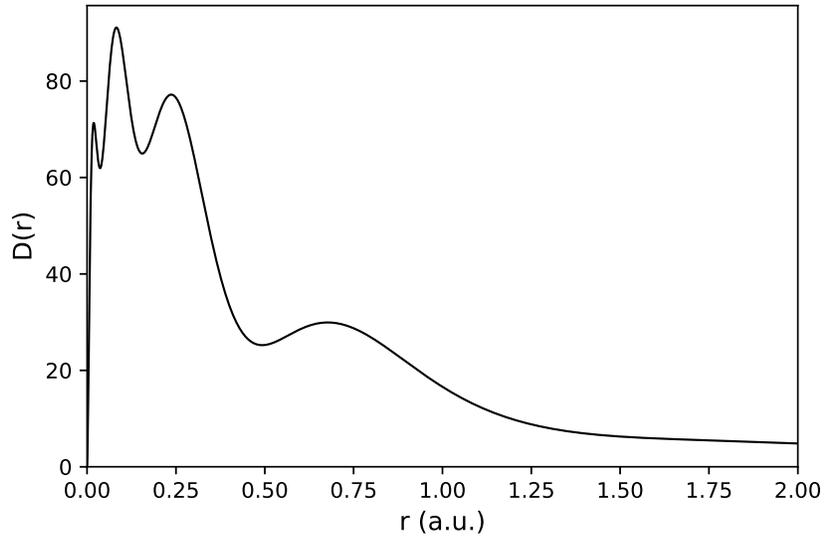}
\caption{}
\label{fig:xe_cg_radialdensityb}
\end{subfigure}

\caption{a) Relativistic radial electronic probability density of Xe atom computed at Hartree--Fock level. b) Change of the radial electronic probability density ($\Delta D(r) =D^\textrm{rel}(r)-D^\textrm{nonrel}(r)$) due to relativistic effects. Obtained at Hartree--Fock level with Dyall-DZ basis~\cite{dyall2016relativistic} using DIRAC~\cite{saue2020dirac} and RHO\_OPS~\cite{rhoops} programs.}
\label{fig:xe_cg_radialdensity}
\end{figure}    

When relativistic effects are taken into account to study the electronic structure of atomic and molecular systems (i.e.\ for electrons traveling at velocities that approach the speed of light ($c$)~\cite{Pyykko.Desclaux.19790gf,Visscher.2002y9,
saue2011relativistic,tecmer2011electronic,samultsev2015relativistic,castro2019four,tiesinga2021relativistic}) the dimensionality of $\Psi$ increases. In addition to the information contained in a non-relativistic wavefunction, a relativistic wavefunction describes also how the direction and magnitude of electron spin depends on the spatial location. Thus, the need for RQs to analyze the changes on the electronic structure becomes a must. Only for reduced quantities and observables (e.g.\ dipole moments), it is possible to directly compare the results of relativistic and non-relativistic calculations. In this line, the effect of relativity in the electronic density has been widely studied (see Refs.~\onlinecite{pyykko1988relativistic,eickerling2007relativistic,fux2011electron,schiffmann2022relativistic}). Indeed, for atomic systems it is well known that relativistic effects lead to a contraction of radial electronic probability density~\cite{schiffmann2022relativistic}, i.e.
\begin{equation}
D(r)=r^2 \int d{\theta}_r d\phi _r n(r,{\theta}_r,\phi _r) \sin (\theta _r) , 
\end{equation}
w.r.t.\ to a non-relativistic approach (see for example Fig.~\ref{fig:xe_cg_radialdensity}).

On the other hand, the effect on many other RQs remains unknown (especially on quantities related to pair probability densities). In this work, we analyze (to the best of our knowledge for the first time) the effect of relativity on the pair probability density (PPD) at the mean-field level employing some auxiliary RQs: a) the (radial) intracule probability density, and b) the (radial) extracule probability density. 

This work is organized as follows: 1) in the theory section we introduce the RQs employed in this work. Then, we introduce the Dirac equation and the Kramers' restricted 4-component spinors that form the set of basis functions employed in relativistic calculations. Next, we present the Coulomb--Gaunt approximation to account for the electron-electron interaction and we introduce the spin-free approximation (to account only for scalar relativistic effects), which is used with just the Coulomb interaction. 2) Then, we analyze the changes due to relativistic effects on the radial intracule probability density (rIPD), where we highlight the effect of relativity on the mean value of the rIPD and on the (integrated) Coulomb electron-electron repulsion energy ($V^{\textrm{C}} _{ee}$). Then, we focus on the contribution to the IPD arising from `opposite-spin' terms; this analysis provides a measure of the relativistic character at the mean-field level. Finally, adjusting the value of speed of light we demonstrate how relativistic effects on the rIPD can also be modified. 3) Following, using the spin-free approximation, we conclude that the scalar relativistic effects play the dominant role in the changes observed on the rIPD due to relativistic effects. And, using the second order Douglas--Kroll--Hess approximation~\cite{nakajima2012douglas,reiher2012relativistic,saue2011relativistic} we have performed post-HF calculations to analyze the changes due to (scalar) relativistic effects when electron correlation effects are also included. 5) Lastly, we present the results obtained for the radial extracule probability density (rEPD), with special attention to the counter-balanced density, which gives the probability of finding the electrons at opposite positions in an (approximately) centro-symmetric system.
\section{Theory}
The RQ of interest in this work is the PPD that can be defined using $\Psi$ as
\begin{align}
n_2({\bf r}_1,{\bf r}_2) = \frac{N(N-1)}{2}\int d{\sigma}_1 d{\sigma}_2 d{\bf x}_3 \dotsi d{\bf x}_N \Psi ^* ({\bf x}_1,{\bf x}_2,{\bf x}_3,\dotsc,{\bf x}_N) \Psi({\bf x}_1,{\bf x}_2,{\bf x}_3,\dotsc,{\bf x}_N), 
\label{Eq:PPD}
\end{align}
where the factor $N(N-1)/2$ accounts for the number of unique electron pairs. The PPD represents the probability of finding electron pairs, where one of the electrons that forms the pair is located at position ${\bf r}_1$ and the other electron is located at position ${\bf r}_2$. This RQ represents a great simplification w.r.t.\ to $\Psi$, but it is still a high-dimensional (6D) function. Thus, in the following, we introduce two 1D RQs that facilitate the analysis\slash{}study of the PPD, which are the rIPD and rEPD.
\subsection{Radial intracule and extracule probability densities.} 
The information contained in the PPD function is commonly analyzed in terms of two (simpler) RQs: a) the intracule probability density (IPD), i.e.
\begin{equation}
P({\bf u})= \int d{\bf r}_1 d{\bf r}_2 \; n_2({\bf r}_1,{\bf r}_2) \delta({\bf u}-{\bf r}_2+{\bf r}_1) 
\label{Eq:IPD}
\end{equation}
with $\delta({\bf R})$ being the 3D Dirac Delta function, and b)
the extracule probability density (EPD), i.e.
\begin{equation}
Q({\bf U})= \int d{\bf r}_1 d{\bf r}_2 \; n_2({\bf r}_1,{\bf r}_2) \delta \left({\bf U}-({\bf r}_2+{\bf r}_1)/2\right).
\label{Eq:EPD}
\end{equation}
While the IPD represents the probability density function for an inter-electronic vector~\cite{debye1915scattering,debye1923zerstreuung,debye1925note,coulson:61ppsl}, the EDP provides a probability density function for the center of mass of an electron pair~\cite{thakkar1981electronic}. Upon integration of the angular coordinates of the IPD one obtains the rIPD that can be written as
\begin{equation}
I(u)= u^2 \int d{\theta}_u d\phi _u P(u,{\theta}_u,\phi _u) \sin (\theta _u) ,
\label{Eq:rIPD}
\end{equation}
which can be easily analyzed and visualized in 2D plots.

Similarly for the rEPD on obtains 
\begin{equation}
E(U)=U^2 \int d {\theta}_U d \phi _U Q(U,{\theta}_U,\phi _U) \sin (\theta _U).
\label{Eq:rEPD}
\end{equation}
Let us remark that the rIPD can actually be measured experimentally via X-ray scattering~\cite{benesch1973density,thakkar:76cpl,thakkar:84pra,watanabe:watanabemp2004}. Furthermore, the Coulomb electron-electron repulsion ($V^{\textrm{C}}_{ee}$) is an explicit functional of the rIPD, i.e.
\begin{equation}
V^{\textrm{C}}_{ee}\left[I\right]=\int_{{0}}^{{\infty}}  du  \frac{I(u)}{u} ,
\end{equation}
which makes the rIPD an interesting RQ for studying the effects that have influence on the electron-electron interaction (e.g.\ electronic correlation effects~\cite{thakkar:76cpl,boyd1988intracule,ugalde1991evaluation,cioslowski1998electron,valderrama2000intracule,gori2008intracule,via2017salient,via2019singling,rodriguez2019coulomb}). But, the rIPD can also be used to interpret molecular interactions~\cite{sarasola1992laplacian,fradera1999topological,fradera2000interpretation,cioslowski1999topology,mercero:17cjc,rodriguez2018electron}. Finally, let us comment that the rIPD as well as the rEPD are commonly normalized to the number of unique electrons pairs (i.e.\ $N(N-1)/2$) like the PPD~\cite{mcweeny:60rmp,mcweeny:89book}. 

In practice, the PPD is usually expressed in a finite (orthonormal) basis for a restricted calculation as  
\begin{align}
n_2({\bf r}_1,{\bf r}_2) &=  \sum _{{P,Q,R,S}} {}^2D_{PQ} ^{RS} \psi _P ^* ({\bf r}_1) \psi _Q ^* ({\bf r}_2) \psi _R  ({\bf r}_1) \psi _S  ({\bf r}_2) \delta _{\sigma _P \sigma _R}\delta _{\sigma _Q \sigma _S},  
\label{Eq:PPD_2RDM}
\end{align}
where $\{ \psi _P ({\bf r})\}$ are the atomic\slash{}molecular spin orbitals; the indices $P$, $Q$, $R$, and $S$ correspond to spin orbitals; $\delta _{\sigma _P \sigma _R}$ ($\delta _{\sigma _Q \sigma _S}$) indicates that the $P$ and $R$ ($Q$ and $S$) spin orbitals have the same spin (i.e.\ $\sigma_P=\sigma_R=\uparrow$ or $\sigma_P=\sigma_R=\downarrow$ and $\sigma_Q=\sigma_S=\uparrow$ or $\sigma_Q=\sigma_S=\downarrow$); and the spin-with matrix elements read as
\begin{equation}
{}^2D_{PQ} ^{RS} = \begin{dcases*}
\frac{\delta_{PR}\delta_{QS}-\delta_{PS}\delta_{QR}}{2} &$P,Q,R,S \in \text{occ}$ \\
0 &otherwise
\end{dcases*}
\label{Eq:2RDMme}
\end{equation}
within the Hartree--Fock (HF) approximation (i.e.\ when $\Psi$ is approximated as a single-determinant wavefunction $\Psi = \Phi _0$, a.k.a.\ the mean-field level wavefunction)~\cite{cohen1976hartree,herbert:03jcp,rodriguez:17pccp_2}. Ugalde et al.~\cite{ugalde1994upper} demonstrated that at HF level the IPD evaluated at the coalescence point for spin-compensated systems~\footnote{Where $n_\uparrow ({\bf r})=n_\downarrow ({\bf r})=n ({\bf r})/2$ with $n_\sigma({\bf r})$ being the electronic density for the spin-channel $\sigma$.} can be written as
\begin{equation}
P({\bf 0})= \frac{1}{4} \int d{\bf r} \; n^2 ({\bf r}), 
\label{Eq:P0_n2}
\end{equation}
which is a consequence of the definition of the ${}^2D_{PQ} ^{RS}$ matrix elements.

\subsection{The relativistic PPD within the Dirac--Hartree--Fock method.}
Relativistic effects are introduced in quantum mechanics by replacing the Schrödinger equation 
with the Dirac equation. The time-independent Dirac equation of a particle subject to potential ${\bf V}({\bf r})$ is written as
\begin{align}
\left[-\textrm{i} c ({\boldsymbol \alpha } \cdot {\boldsymbol \nabla} _{{\bf r}}) +c^2  m {\boldsymbol \beta}+ {\bf V}({\bf r}) {\bf I}_4\right] {\boldsymbol\psi} _P ({\bf r}) = E_P {\boldsymbol \psi} _P({\bf r}) ,
\label{QEDH}
\end{align}
where ${\bf I}_4$ is the $4 \times 4$ unit matrix, $\textrm{i} = \sqrt{-1}$, $c=137.036$\;a.u., $m=1$\;a.u.\ is the electron mass,
\begin{align}
{\boldsymbol \alpha }&= ({\boldsymbol\alpha} _{x},{\boldsymbol\alpha} _{y},{\boldsymbol\alpha} _{z}) \nonumber \\
&=\left({
\begin{pmatrix}
{\boldsymbol 0}_2 & {\boldsymbol\sigma} _{x} \\
{\boldsymbol\sigma} _{x} & {\boldsymbol 0}_2
\end{pmatrix}
},
{
\begin{pmatrix}
{\boldsymbol 0}_2 & {\boldsymbol\sigma} _{y} \\
{\boldsymbol\sigma} _{y} & {\boldsymbol 0}_2
\end{pmatrix}
},
{
\begin{pmatrix}
{\boldsymbol 0}_2 & {\boldsymbol\sigma} _{z} \\
{\boldsymbol\sigma} _{z} & {\boldsymbol 0}_2 
\end{pmatrix}
}\right),
\label{alphamat}
\end{align}
${\boldsymbol 0}_2$ is the 2$\times$2 null matrix, ${\boldsymbol \sigma} _{x}$, ${\boldsymbol \sigma} _{y}$, ${\boldsymbol \sigma} _{z}$ are the Pauli matrices;  
\begin{equation}
{\boldsymbol \beta } = 
\begin{pmatrix}
{\bf I}_2 &  {\boldsymbol 0}_2  \\
{\boldsymbol 0}_2 & - {\bf I}_2 
\end{pmatrix},
\end{equation}
${\bf I}_2$ is the 2$\times$2 unit matrix. Solutions of the Dirac equation (${\boldsymbol \psi}_P$) are 4-component-spinor orbitals (that can also be written as 2-component-spinors, one for the so-called large component ${\boldsymbol \psi} ^L _{P} ({\bf r})$ and another ${\boldsymbol \psi} ^S _{P} ({\bf r})$ for the small component)  
\begin{equation}
{\boldsymbol \psi} _P ({\bf r}) =    
\begin{pmatrix}
{\boldsymbol \psi} ^L _{P} ({\bf r})  \\
{\boldsymbol \psi} ^S _{P} ({\bf r})  \\
\end{pmatrix}
=
\begin{pmatrix}
\chi _{P,1} ({\bf r})  \\
\chi _{P,2} ({\bf r})  \\
\chi _{P,3} ({\bf r})  \\
\chi _{P,4} ({\bf r}) 
\end{pmatrix},
\end{equation}
whose conjugate-transpose form reads 
\begin{align}
{\boldsymbol \psi} ^\dagger _P ({\bf r}) &=    
\begin{pmatrix}
\chi ^{*} _{P,1} ({\bf r}) & \chi _{P,2} ^{*} ({\bf r}) & \chi _{P,3} ^{*} ({\bf r}) & \chi _{P,4} ^{*} ({\bf r}) 
\end{pmatrix}.
\end{align}
These 4-component-spinor orbitals form an orthonormal basis set (i.e.\ $\int d{\bf r} {\boldsymbol \psi}^\dagger _P ({\bf r}) {\boldsymbol \psi} _Q ({\bf r})=\delta _{PQ}$)~\footnote{The $P$ and $Q$ indices label positive and negative energy 4-component spinors.}. Let us remark that the scalar orbitals $\chi _{P,\mu}({\bf r})$ do not contain any information about spin (this information is given by the component of the spinor where the scalar orbital is located). Moreover, the 4-component spinors describe electronic and positronic states, with the latter best interpreted in the framework of quantum electrodynamics\cite{liu2014advances,liu2017handbook,salman2022quantum,toulouse2021relativistic,sunaga20224}. In this work, we will not consider such states explicitly and apply the no-pair approximation~\cite{sucher1980foundations,mittleman1981theory} in which only the electronic states are employed to build $\Psi$. 

The number of components of an $N$-electron wavefunction $\Psi$ constructed from 4-component spinors is formally 4$^N$. This is an increase compared to the non-relativistic wavefunction that is formally 2$^N$, but with the different spin-components not being coupled by the Hamiltonian if magnetic fields are absent. In non-relativistic theory it is therefore possible to integrate out the spin and work with spatial wavefunctions, so-called configuration state functions. In relativistic calculations, the Hamiltonian does couple in principle all components of the wavefunction which makes such a simplification impossible. 
The relativistic PPD is evaluated as
\begin{align}
\label{Eq:rel_PPD_2RDM}
n_2&({\bf r}_1,{\bf r}_2) = \sum _{\mathclap{P,Q,R,S}} {}^2D_{PQ} ^{RS} \bigl({\boldsymbol \psi} _P ^\dagger ({\bf r}_1) \otimes {\boldsymbol \psi} _Q ^\dagger ({\bf r}_2) \bigl) \bigl({\boldsymbol \psi} _R  ({\bf r}_1) \otimes {\boldsymbol \psi} _S  ({\bf r}_2)\bigl), 
\end{align}
with $\otimes$ denoting the tensor product~\footnote{Notice that the indices $P$, $Q$, $R$, $S$ refer to occupied 4-component spinors in the relativistic context.}.
Note that since also for the relativistic PPD we integrate out spin,
we can compare the relativistic PPD directly with the non-relativistic PPD.

To facilitate our analysis and comparison with the restricted non-relativistic result, in this work we adopt a Kramers' restricted formalism to construct the 4-component spinors. To indicate the difference with the unrestricted spinor formalism we label such pairs with lowercase symbols, in which we distinguish between the two Kramers' paired spinors by labeling one of them with an overbar. The 4-component spinors that form a Kramers' pair ($p,\bar{p}$) transform as $\widehat{\mathcal{K}} {{\boldsymbol \psi}} _p ={{\boldsymbol \psi}} _{\bar{p}}$ and $\widehat{\mathcal{K}} {{\boldsymbol \psi}} _{\bar{p}} =-{{\boldsymbol \psi}} _p $ upon application of the time-reversal operator. The operator $\widehat{\mathcal{K}}$ is defined as
\begin{equation}
\widehat{\mathcal{K}} =-\textrm{i} 
{
\begin{pmatrix}
{\boldsymbol \sigma} _y & {\boldsymbol 0}_2 \\
{\boldsymbol 0}_2 & {\boldsymbol \sigma} _y
\end{pmatrix}
} \widehat{\mathcal{K}}_0, 
\label{Kramers}
\end{equation}
where $\widehat{\mathcal{K}}_0$ is the complex conjugation operator. It is important to realize that the choice of the set of unbarred spinors $p$ labels that comprise half of the Hilbert space spanned by the spinors is arbitrary. The only constraint is that by operation on this set of spinors by $\widehat{\mathcal{K}}$ one should obtain a set of spinors that spans the other half. This arbitrariness makes it difficult to construct algorithms that mimic non-relativistic restricted algorithms\cite{Visscher.Lee.1995,Fleig.2008, Komorovsky.2016}. Only when the system possess additional symmetries, such as a 2-fold rotation axis or a mirror plane, it is possible to unambiguously define a set of unbarred and barred spinors\cite{Visscher.1996}. In other cases, one may resort to quaternion algebra to ensure proper pairing of the spinors.\cite{Saue.Jensen.1999ttz,Senjean.2021}  In the non-relativistic limit, pairing of spin orbitals is also arbitrary, but there exist the natural choice to select the unbarred spinors as $\alpha$ spin-orbitals, so the barred spinors then correspond to $\beta$-spin orbitals. Given this limit, imposing time-reversal symmetry on the spinors is denoted as a restricted formalism. When the relativistic $\Psi$ is approximated as a single-determinant wavefunction, the approximation is known as the Kramers' restricted Dirac--Hartree--Fock (DHF) method~\cite{hafner1980kramers,sun2021efficient}. The DHF method leads to a mean-field treatment of the electron-electron interaction (as in the non-relativistic case), where an equation similar to Eq.~\eqref{QEDH} is used to obtain the optimal 4-component spinors (replacing the ${\bf V} ({\bf r})$ by an effective potential ${\bf V}_{\textrm{eff}} ({\bf r})$ that also accounts for the Hartree and exchange contributions, see Ref.~\cite{szabo2012modern} for more details). For this method, the `spin-with' matrix elements\footnote{The ones in which spin integration is always over fully barred or  fully unbarred spinor products.} ${}^2D_{PQ} ^{RS}$ that enter Eq.~\eqref{Eq:rel_PPD_2RDM} are the same as in the non-relativistic context (i.e.\ they are given by Eq.~\eqref{Eq:2RDMme}). However, in the relativistic case also the `exchange of opposite-spin terms' (i.e.\ where spin integration is over barred and unbarred spinors) contributes to the PPD, (r)IPD and (r)EPD as spinors have then both $\alpha$ and $\beta$ character. This `exchange of opposite-spin terms' contribution is not present in the non-relativistic context because the spin integration then produces a zero value. 

\subsection{The electron-electron interaction.}
In the relativistic context the Coulomb electron-electron interaction ($W^{\textrm{C}}({\bf r}_1,{\bf r}_2)=|{{\bf r}_2-{\bf r}_1}|^{-1}$) is only a low order approximation to the actual interaction (that should be Lorentz invariant and mediated by transferring photons). Therefore, an easy manner to include a correction is with the magnetic Gaunt interaction~\cite{gaunt1929iv} which combined with the Coulomb interaction is called the Coulomb--Gaunt (CG) interaction, i.e.
\begin{equation}
W^{\textrm{CG}}({\bf r}_1,{\bf r}_2)= \frac{\left[1 -{\boldsymbol \alpha} \odot {\boldsymbol \alpha}\right]}{|{{\bf r}_2-{\bf r}_1}|} , 
\end{equation}
where \(\boldsymbol{\alpha} \odot \boldsymbol{\alpha} = \boldsymbol{\alpha}_x \otimes \boldsymbol{\alpha}_x + \boldsymbol{\alpha}_y \otimes \boldsymbol{\alpha}_y + \boldsymbol{\alpha}_z \otimes \boldsymbol{\alpha}_z\). Though this approximation is not Lorentz invariant, it is an improvement over the bare Coulomb interaction~\cite{dyall2007introduction}. 

\subsection{Scalar relativistic effects and electronic correlation.}
To analyze the role of the scalar relativistic effects (i.e.\ neglecting spin-orbit coupling effects), we have employed the spin-free (SF) approximation, which is derived from a modified Dirac equation~\cite{dyall1994exact} by exactly splitting this equation into spin-free and spin-dependent contributions. If Kramers' pairing is achieved by quaternion algebra, as it is in the DIRAC program\cite{saue2020dirac} that we use in this work, such a spin-free formalism can be implemented by keeping only the real part of the quaternion Fock matrix\cite{Visscher.2000}. To be consistent and neglect also the 2-electron spin-other-orbit contribution contained in the Gaunt interaction\cite{Saue.1996}, we have used this SF approximation including only the bare Coulomb interaction, i.e.
\begin{equation}
W^{\textrm{C}}({\bf r}_1,{\bf r}_2)= \frac{1}{|{{\bf r}_2-{\bf r}_1}|}.
\end{equation}
This approach then accounts only for the (one-body) scalar relativistic effects and we ensure that all the relativistic effects that modify the rIPD are purely scalar one-body ones. The SF Dirac approximation provides results without spin-orbit coupling effects; thus, it does not split the shells and the shell structure is qualitatively the same as in the non-relativistic case. An alternative, more approximate treatment of scalar relativistic effects is given by the second-order Douglas--Kroll--Hess approximation~\cite{douglas:53jcp,hess1985applicability,hess1986relativistic,jansen1989revision} (DKH), which we have employed in combination with the configuration interaction including singles and doubles excited determinants (CISD)~\cite{helgaker:00book,szabo2012modern} approximation, i.e.\ approximating the wavefunctions as
\begin{equation}
\Psi ^{\textrm{CISD}} = \Phi_0 + \sum _{i,a} C_{i,a} \Phi_{i\rightarrow a}+ \sum _{\mathclap{i,j,a,b}} C_{ia,jb} \Phi_{i\rightarrow a} ^{j\rightarrow b} ,
\label{eq:cisd}
\end{equation}
to analyze the changes in the rIPD when electron-correlation effects are also considered. Let us mention that in Eq.~\eqref{eq:cisd} the $i$ and $j$ ($a$ and $b$) indices refer to states that are (un)occupied at the HF level and that the lowercase indicates that the CISD is spin-restricted\footnote{The unoccupied states correspond to the virtual states.}.

\section{Results and discussions.}
In this work we have used the DIRAC~\cite{saue2020dirac} program and the Dyall-DZ~\cite{dyall2016relativistic} basis set to perform DHF calculations with the $W^{\textrm{CG}}({\bf r}_1,{\bf r}_2)$ interaction. This basis was also employed in PSI4~\cite{smith2020psi4} program to run the corresponding non-relativistic Restricted Hartree--Fock calculations. Furthermore, this basis was used in Gaussian 03~\cite{g03c02} package to run calculations including scalar relativistic effects by means of the DKH and including electronic correlation effects by using the CISD approximation. The correlated 2-RDMs for the CISD wavefunctions were computed with the DMn code~\cite{dmn}. IPDs and EPDs were computed using a modified version of RHO2\_OPS~\cite{rodriguezrho2ops} code that uses the algorithm of Cioslowski and Liu~\cite{cioslowski:96jcpintra}. The integral $\int d{\bf r} \; n^2({\bf r})$ was computed with a modified version of RHO\_OPS~\cite{rodriguezrhoops}. Python scripts were used for plotting the data including the \verb+make_interp_spline+ option from the \verb+scipy.interpolate+ library to construct cubic interpolations for certain rough plots. For diatomic systems (i.e.\ H$_2$, LiH, NaH, and RbH) the heaviest element was placed at the origin of coordinates for the computation of the rEPD and the counter-balanced density; the experimental equilibrium geometries were employed~\cite{johnson2013nist,okorie2020improved}. Finally, let us comment that the PSI4, Gaussian, and DIRAC programs employ Gaussian basis sets to represent the atomic orbitals. The use of Gaussian basis sets leads to a poor description of the electron-electron cusp condition~\cite{kato:57cpam,helgaker:00book}. Nevertheless, we may consider that the poor description of that cusp is similar in our non-relativistic and relativistic calculations; thus, we could expect that some error cancellation may take place in the description of $P({\bf 0})$. Also, it is worth to mention that in the DIRAC program we have employed the default finite size distribution for the nuclei~\cite{visscher1997dirac}, while in PSI4 and Gaussian 03 we used a point charge model for the nuclei. Our results indicate (see below) that using a different model for the nuclei does not have a large effect on the differences in the rIPD and rEPD due to relativistic effects.

\begin{figure}[H]
\centering
\begin{subfigure}[t]{\textwidth}
\centering
\includegraphics[scale=0.8]{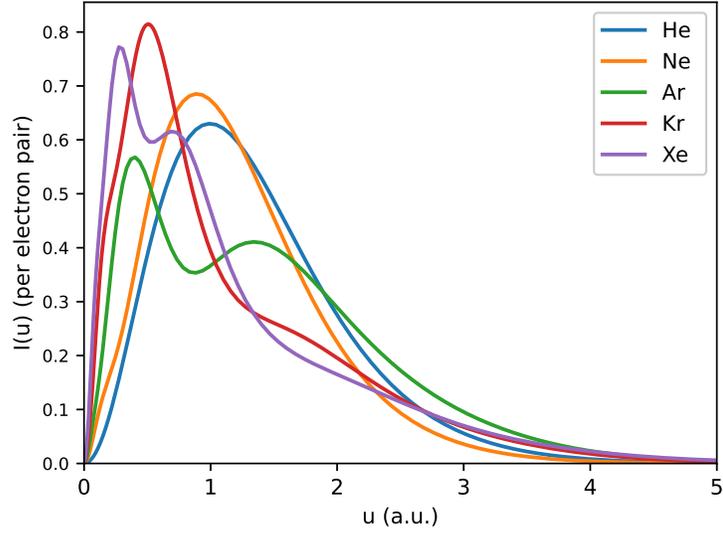}
\caption{}
\label{fig:ngas_cg_intraa}
\end{subfigure}
\hfill
\begin{subfigure}[t]{\textwidth}
\centering
\includegraphics[scale=0.8]{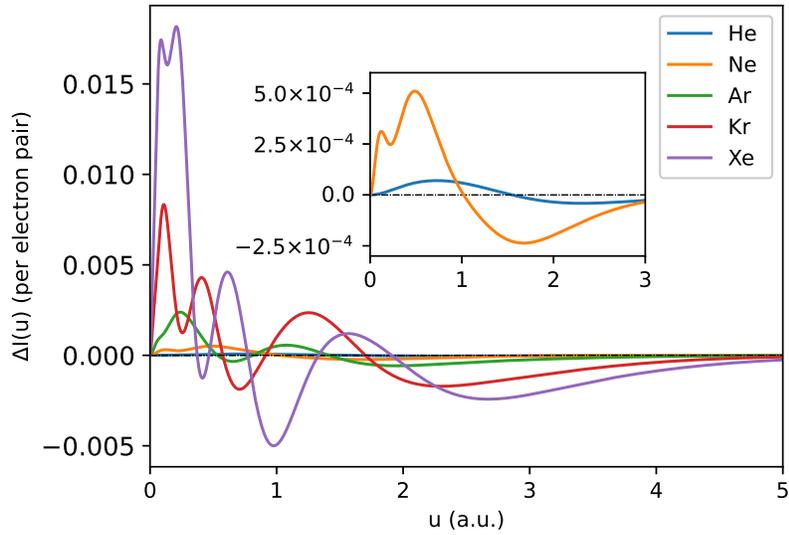}
\caption{}
\label{fig:ngas_cg_intrab}
\end{subfigure}

\caption{a) Relativistic rIPD for the noble gases (from He to Xe). b)Change on the rIPD ($\Delta I(u) = I^{\textrm{rel}}(u)- I^{\textrm{nonrel}}(u)$) for the noble gases (from He to Xe) due to relativistic effects. Results obtained with Dyall-DZ basis~\cite{dyall2016relativistic}.}
\label{fig:ngas_cg_intra}
\end{figure}

\subsection{Relativistic effects on the rIPD.}
\subsubsection{Contraction of the inter-electronic distance.}
We have computed the relativistic rIPD using the Dyall-DZ basis set~\cite{dyall2016relativistic} at DHF level and show it in Fig.~\ref{fig:ngas_cg_intraa} rescaled per electron pair to be able to compare their shapes. Our results indicate that for the He, Ne, and Kr atoms only one peak in the rIPD is obtained. On the other hand, the rIPD of Ar and Xe atoms presents two peaks. In all cases, the peak(s) displace towards the origin when the atomic number ($Z$) increases, which is a consequence of the increase in the number of possible electron pairs that can be formed near the progressively stronger electric attractive field produced by the nuclei.

\newpage
\vspace*{-3cm}\begin{table}[H]
\resizebox{6.5cm}{11.5cm}{
    \caption{$\langle u \rangle$  of the radial intracule probability density (rIPD), $\sigma _u ^2 = \langle u ^2 \rangle-\langle u \rangle^2$ of the rIPD, $\langle U \rangle$ of the radial extracule probability density (rEPD), and $\sigma _U ^2 = \langle U ^2 \rangle-\langle U \rangle^2$ of the rEPD computed at Hartree--Fock (HF) and Dirac--Hartree--Fock (DHF) levels using the Dyall-DZ basis~\cite{dyall2016relativistic}}.
    \label{tab:index}
    \begin{tabular}{cccccc}\hline
    \multicolumn{2}{c}{System} & $\langle u \rangle$\;a.u.\ & $\sigma _u ^2$ & $\langle U \rangle$\;a.u.\ & $\sigma _U ^2$\\ \hline
    \multicolumn{6}{c}{Alkali metals} \\ 
    \multirow{2}{*}{Li} & (HF)  & 2.8951 & 4.0175  & 1.4476  & 1.0044 \\ 
    &(DHF) & 2.8949 & 4.0172 & 1.4476 & 1.0044 \\ 
    \multirow{2}{*}{Na} & (HF)  & 1.6173 & 2.3559  & 0.7895  & 0.5965 \\ 
    &(DHF) & 1.6159 & 2.3452 & 0.7895 & 0.5965 \\ 
    \multirow{2}{*}{K} & (HF)   & 1.6894 & 2.5683  & 0.8307  & 0.6419 \\ 
    &(DHF)  & 1.6856 & 2.5489 & 0.8307 & 0.6419 \\ 
    \multirow{2}{*}{Rb} & (HF)  & 1.3273 & 1.9423  & 0.6563  & 0.4845 \\ 
    &(DHF) & 1.3158 & 1.8990 & 0.6563 & 0.4845 \\ 
    \multicolumn{6}{c}{Alkaline-Earth metals} \\
    \multirow{2}{*}{Be} & (HF)  & 2.5185 & 2.2921 & 1.2593 & 0.5730 \\ 
    &(DHF) & 2.5164 & 2.2816 & 1.2582 & 0.5704 \\ 
    \multirow{2}{*}{Mg} & (HF)  & 1.6787 & 2.1333 & 0.8250 & 0.5418 \\ 
    &(DHF) & 1.6763 & 2.1229 & 0.8239 & 0.5391 \\ 
    \multirow{2}{*}{Ca} & (HF)  & 1.7587 & 2.5947 & 0.8675 & 0.6509 \\ 
    &(DHF) & 1.7535 & 2.5714 & 0.8650 & 0.6450 \\ 
    \multirow{2}{*}{Sr} & (HF)  & 1.3852 & 2.0736 & 0.6859 & 0.5183 \\ 
    &(DHF) & 1.3738 & 2.0422 & 0.6803 & 0.5103 \\ 
    \multicolumn{6}{c}{Transition metals II B} \\
    \multirow{2}{*}{Zn} & (HF)  & 1.1732 & 0.9568 & 0.5789 & 0.2405 \\ 
    &(DHF) & 1.1675 & 0.9359 & 0.5761 & 0.2353 \\ 
    \multirow{2}{*}{Cd} & (HF)  & 1.1454 & 0.9268 & 0.5675 & 0.2317 \\ 
    &(DHF) & 1.1326 & 0.9078 & 0.5610 & 0.2269 \\ 
    \multicolumn{6}{c}{Halogens} \\
    \multirow{2}{*}{F} & (HF)  & 1.3515 & 0.5358 & 0.6448 & 0.1268 \\ 
   &(DHF) & 1.3511 & 0.5357 & 0.6446 & 0.1268 \\ 
    \multirow{2}{*}{Cl} & (HF)  & 1.5000 & 1.0310 & 0.7321 & 0.2487 \\ 
    &(DHF) & 1.4983 & 1.0303 & 0.7312 & 0.2486 \\ 
    \multirow{2}{*}{Br} & (HF)  & 1.1910 & 0.8923 & 0.5873 & 0.2196 \\ 
   &(DHF) & 1.1846 & 0.8826 & 0.5842 & 0.2173 \\ 
    \multirow{2}{*}{I}  & (HF)  & 1.1829  & 0.9848 & 0.5857 & 0.2433 \\ 
   &(DHF) & 1.1701 & 0.9687 & 0.5794 & 0.2394 \\ 
    \multicolumn{6}{c}{Noble gases} \\
    \multirow{2}{*}{He} & (HF)  & 1.3613 & 0.5107 & 0.6806 & 0.1277 \\ 
   &(DHF) & 1.3612 & 0.5105 & 0.6806 & 0.1276 \\ 
    \multirow{2}{*}{Ne} & (HF)  & 1.2263 & 0.4351 & 0.5867 & 0.1040 \\ 
   &(DHF) & 1.2259 & 0.4351 & 0.5866 & 0.1040 \\ 
    \multirow{2}{*}{Ar} & (HF)  & 1.4316 & 0.8764 & 0.6989 & 0.2117 \\ 
   &(DHF) & 1.4298 & 0.8759 & 0.6981 & 0.2116 \\ 
    \multirow{2}{*}{Kr} & (HF)  & 1.1723 & 0.8191 & 0.5782 & 0.2015 \\ 
   &(DHF) & 1.1661 & 0.8110 & 0.5751 & 0.1996 \\ 
    \multirow{2}{*}{Xe} & (HF)  & 1.1753 & 0.9350 & 0.5819 & 0.2309 \\ 
   &(DHF) & 1.1622 & 0.9192 & 0.5756 & 0.2271 \\
    \multicolumn{6}{c}{hydrides} \\
    \multirow{2}{*}{H$_2$} & (HF) & 2.0429 & 0.9977 & 1.0214 & 0.2494 \\ 
   &(DHF) & 2.0429 & 0.9979 & 1.2128 & 0.3121 \\ 
    \multirow{2}{*}{LiH} & (HF)   & 2.9182 & 2.2536 & 1.5051 & 0.7525 \\ 
   &(DHF) & 2.9171 & 2.2518 & 1.7229 & 0.9446 \\ 
    \multirow{2}{*}{NaH} & (HF)   & 1.9442 & 2.5378 & 0.9624 & 0.6473 \\ 
   &(DHF) & 1.9436 & 2.5356 & 0.9842 & 0.7383 \\ 
    \multirow{2}{*}{KH} & (HF)    & 1.9184 & 2.6985 & 0.9513 & 0.6716 \\ 
   &(DHF) & 1.9175 & 2.6969 & 0.9578 &  0.7276 \\ 
   \multirow{2}{*}{RbH} & (HF)    & 1.4688 & 2.0939 & 0.7296 & 0.5224 \\ 
   &(DHF) & 1.4608 & 2.0959 & 0.7269 & 0.5401 \\ \hline 
    \end{tabular}
}
\end{table}

Let us focus on the comparison between the relativistic and the non-relativistic rIPD that is given in Fig.~\ref{fig:ngas_cg_intrab}, where we have plotted the difference between the relativistic and non-relativistic rIPDs (i.e.\ $\Delta I(u) = I^{\textrm{rel}}(u)- I^{\textrm{nonrel}}(u)$). Let us mention that with the usual Dunning\cite{dunning:89jcp} cc-pVDZ and cc-pVTZ basis sets we obtained similar results in all cases when these basis were available. But, the Dyall-DZ basis sets were preferred in this work for two reasons: a) they are available for heavy elements, and b) the sets are especially designed for treating relativistic effects with a rather compact basis~\footnote{The computation of the rIPD and rEPD is very expensive from the computational perspective; thus, with the Dyall-DZ basis it was still possible to calculate it for a reasonable computational cost.}. Our results suggest that for small $u$ values the relativistic rIPD is larger than its non-relativistic counterpart (i.e.\ there is an increase of probability for the electrons to be closer to each other). This result is consistent with the contraction of the radial electronic density ($D(r)$), cf.\ Fig.~\ref{fig:xe_cg_radialdensityb}, which indicates that the increase of $D(r)$ near the nucleus also increases the number of electron pairs in this region. Lastly, as one could anticipate, this effect increases with $Z$ (being almost negligible for light atoms like He and Ne).     

\begin{figure}[H]
\centering
\begin{subfigure}[t]{\textwidth}
\centering
\includegraphics[scale=0.8]{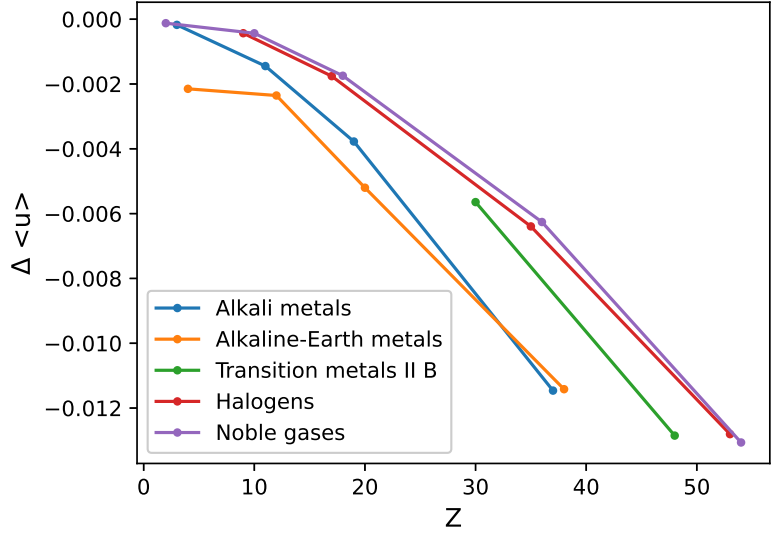}
\caption{}
\label{fig:all_cg_MeanUVeeCa}
\end{subfigure}
\hfill
\begin{subfigure}[t]{\textwidth}
\centering
\includegraphics[scale=0.8]{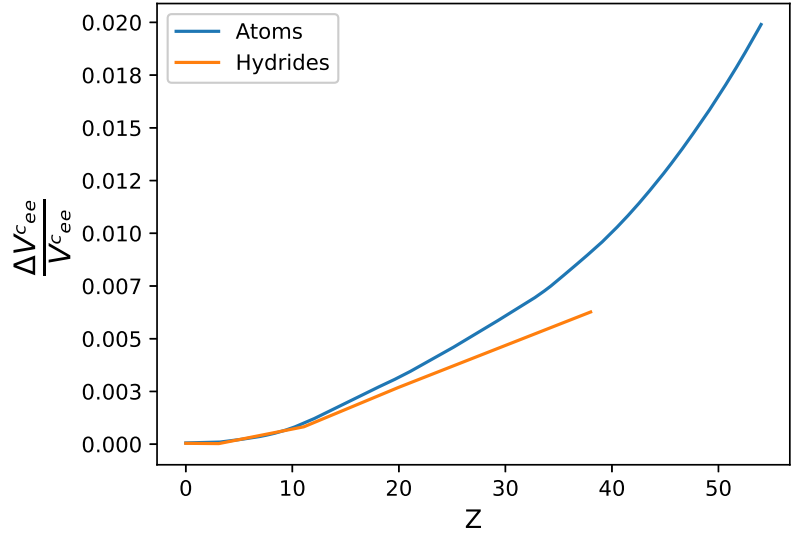}
\caption{}
\label{fig:all_cg_MeanUVeeCb}
\end{subfigure}

\caption{a) Change on the mean value of the rIPD ($\Delta \langle u \rangle = \langle u \rangle^{\textrm{rel}}- \langle u \rangle ^{\textrm{nonrel}}$) due to relativistic effects. {b) Relative change on the integrated Coulomb electron-electron repulsion energy $\Delta V^{\textrm{C}}_{ee}/V^{\textrm{C}}_{ee}=(V^{\textrm{C,rel}}_{ee}-V^{\textrm{C,nonrel}}_{ee})/{V^{\textrm{C,rel}}_{ee}}$ due to relativistic effects.} Results obtained with Dyall-DZ basis~\cite{dyall2016relativistic}. For the hydrides the $Z$ value corresponds to the atomic number of the heaviest element.}
\label{fig:all_cg_MeanUVeeC}
\end{figure}

The increased probability for shorter inter-electronic distances, $u$, is reflected in a reduction of the mean value of the radial distribution $\langle u \rangle = \int d u I(u)u$, which is shown in Fig.~\ref{fig:all_cg_MeanUVeeCa} as $\Delta \langle u \rangle= \langle u \rangle ^{\textrm{rel}}- \langle u \rangle^{\textrm{nonrel}} $ w.r.t.\ the atomic number for all the atomic systems studied in this work (the values of $ \langle u \rangle^{\textrm{(non)rel}}$ are collected in Table~\ref{tab:index}). Our results show a dependence of $\Delta \langle u \rangle$ with $Z$ for each group of the periodic table, where the largest deviation is obtained for Xe atom (i.e.\ the heaviest element employed in this study). Let us remark that for Alkali metals and Halogens a restricted open-shell formalism~\cite{Thyssen.20029nt} was employed. The contraction of the inter-electronic distance induces an increase of the Coulomb repulsion among the electrons as we can see in Fig.~\ref{fig:all_cg_MeanUVeeCb}, {where we have plotted the relative change in the electron-electron energy (i.e.\ \(\Delta V^{\textrm{C}}_{ee}/V^{\textrm{C}}_{ee}=(V^{\textrm{C,rel}}_{ee}-V^{\textrm{C,nonrel}}_{ee})/{V^{\textrm{C,rel}}_{ee}}\)) for all the systems studied in this work\footnote{The relative error is plotted because the absolute change (i.e.\ $V^{\textrm{C,rel}}_{ee}-V^{\textrm{C,nonrel}}_{ee}$) also accounts for the increase in the number of electron when $Z$ gets larger.}}. The $\Delta V^{\textrm{C}}_{ee}/V^{\textrm{C}}_{ee}$ increases rapidly w.r.t.\ $Z$, where a difference of $\sim$0.02 was obtained for Xe.

In Table~\ref{tab:index} we have also collected the variance of the rIPD ($\sigma _u ^2 = \langle u^2 \rangle - \langle u \rangle ^2$) for all the systems studied. Our results indicate that there is a small reduction of the variance, which suggests that the spread of the rIPD is also reduced due to relativistic effects at the mean-field level. 

We have furthermore analyzed the inter- and intra-shell contributions to $\Delta I(u)$ in Xe atom (see the Appendix~\ref{appexA} for more details). Our results suggest that all shells are involved in the change of the rIPD. Nonetheless, it is clear that the major changes on the rIPD (due to relativistic effects) are of inter-shell type. Furthermore, we have noticed that the first peak observed on $\Delta I(u)$ at very short inter-electronic distances is mostly due to changes involving the 1s shell (of both types i.e.\ intra- and inter-shell contributions), where the electron-pairs are on average at shorter inter-electronic distances than in the rest of the shells.

\subsubsection{The effect of the exchange of `opposite-spin' terms.}
\begin{figure}[H]
\includegraphics[scale=0.8]{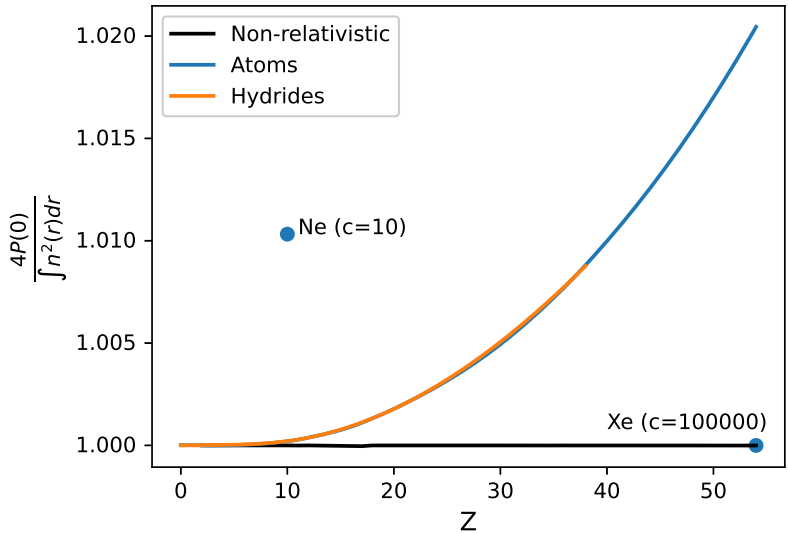}
\caption{Effect of the exchange of `opposite-spin' terms at the intracule coalescence point. Results obtained with Dyall-DZ basis~\cite{dyall2016relativistic}. For the hydrides the $Z$ value corresponds to the atomic number of the heaviest element. The dots correspond to the deviations obtained with Ne and Xe atoms varying the value of the speed of light.}
\label{fig:all_cg_dev1}
\end{figure}

In non-relativistic theory, the deviation from unity of the ratio
\begin{equation}
\frac{4 P({\bf 0})}{\int d{\bf r} n^2 ({\bf r})}
\label{ratio_equation}
\end{equation} 
has been used by Ugalde et al.~\cite{ugalde1994upper} to characterize the presence of electronic correlation effects, because this ratio is only equal to 1 at the R(O)HF level. Let us recall, however, that in the derivation of Eq.~\eqref{Eq:P0_n2} a non-relativistic approach was employed, where the exchange of `opposite-spin' elements is not present. This implies that in the relativistic case, also without accounting for electron correlation effects, a deviation from unity in equation (\ref{ratio_equation}) is expected, so can be used to indicate the relativistic character for systems containing more than two electrons (see also appendix B in Ref.~\cite{paquier2020relativistic}). In Fig.~\ref{fig:all_cg_dev1} we have plotted this ratio versus $Z$. { Let us first remark that the ratio given by Eq.~\eqref{Eq:P0_n2} decreased (i.e.\ is $\leq 1$) when electronic-correlation effects were considered in non-relativistic calculations~\cite{ugalde1994upper}. On the contrary, the results in Fig.~\ref{fig:all_cg_dev1} suggest that the (integrated) on-top pair density\footnote{Recall that the on-top pair density is defined as $\Pi({\bf r})=n_2({\bf r},{\bf r})$.} increases when relativistic effects are taken into account (we have collected in Appendix~\ref{appexB} the (integrated) on-top pair density values that confirm this observation).}

{ From Fig.~\ref{fig:all_cg_dev1}} we observe that the deviation from 1 in the relativistic case increases rapidly with $Z$.
To verify this suspicion, we have adjusted the value of the speed of light ($c$) to modulate the relativistic effects~\footnote{This is a common practice in the community to tune relativistic effects (see Ref.~\onlinecite{ebert1996})}. First, we have approached to the non-relativistic limit by increasing the value of $c$ for Xe atom. Next, we went in the other direction by reducing $c$ to enhance the relativistic effects in Ne atom. Our results confirm that using a value of $c=10$\;a.u.\ for the Ne atom the relativistic effects are enhanced the on the (r)IPD, where a much larger reduction of the mean inter-electronic distance is obtained (See the Appendix~\ref{appexC} for more details). Moreover, the ratio $4P({\bf 0})/\int d{\bf r} n^2 ({\bf r})=1.010322$ for Ne atom when a value of $c=10$\;a.u.\, is employed, which suggests an increase in the contribution of the `opposite-spin' terms (i.e.\ of the 4-component relativistic character). In the case of Xe atom, an increase of the value of the speed of light to $c$=100\,000\;a.u.\ is able to switch off almost all relativistic effects. Also, the effect of `opposite-spin' is reduced since the ratio is again close to the non-relativistic result, which indicates a correct approach of the 4-component orbitals (${\boldsymbol \psi}$) to their non-relativistic counterparts (${\psi}$) as should be guaranteed by the kinetic balance condition that is used to define the small component basis set\cite{Stanton.Havriliak.1984zc}.

\subsubsection{The SF rIPD}
\begin{figure}[H]
\includegraphics[scale=0.8]{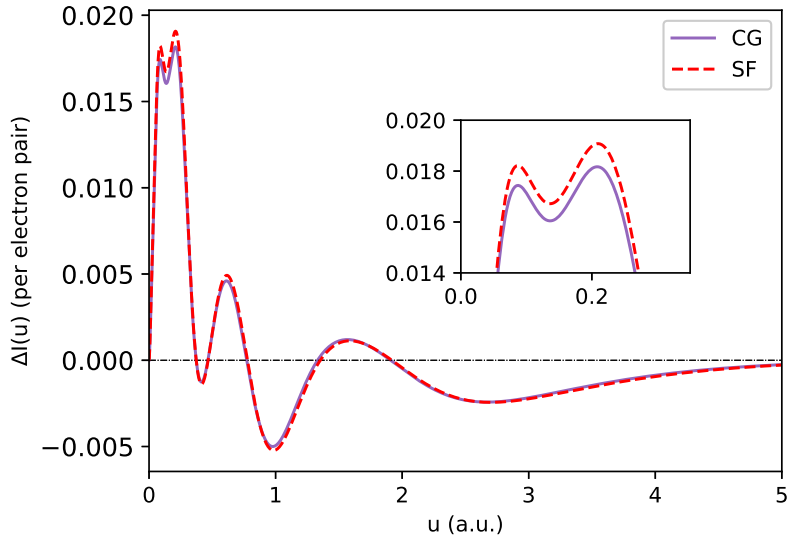}
\caption{Change on the rIPD due to relativistic effects for Xe atom computed with and without spin-orbit coupling effects ($\Delta I(u) = I^{\textrm{rel}}(u)- I^{\textrm{nonrel}}(u)$). Results obtained with Dyall-DZ basis~\cite{dyall2016relativistic}.}
\label{fig:xe_cg_sf_intra}
\end{figure}

The analysis of change in the rIPD w.r.t.\ the non-relativistic one when the SF approximation is employed shows that the reduction of the inter-electronic distance does not depend much on the spin-orbit coupling effects (see Fig.~\ref{fig:xe_cg_sf_intra}). Actually, the contraction of the inter-electronic distance is ({slightly}) more pronounced with the SF than in the previous case (see Fig.~\ref{fig:xe_cg_sf_intra})\footnote{{This observation will be further analyzed in future works because it requires additional functionality in our implementation: a) the 16$\times$16 rIPD, b) a systematic manner to turn off two-electron spin-orbit coupling effects, and, c) the analysis of the role of the self-consistent procedure in the spinor (orbital) optimization when these terms are modified.}}.
This result suggests that the change in the rIPD is mostly due to the scalar relativistic effects (similar results were obtained when the Zeroth-Order Regular Approximation (ZORA) approximation~\cite{faas1995zora,saue2011relativistic} was employed). Also, the SF approximation produces the contraction of the electronic density; thus, we can argue that the contraction in the inter-electronic distance is driven by the contraction in the radial electronic density. Hence, the change on the rIPD can be considered as mainly a density driven effect. Finally, let us comment that  using different approximations to the electron-electron interaction very similar $\Delta I (u)$ plots are obtained, which also points out the fundamental role played by the scalar relativistic effects on the $\Delta I(u)$ changes (See the Appendix~\ref{appexC} for more details).

\begin{figure}[H]
\centering
\includegraphics[scale=0.8]{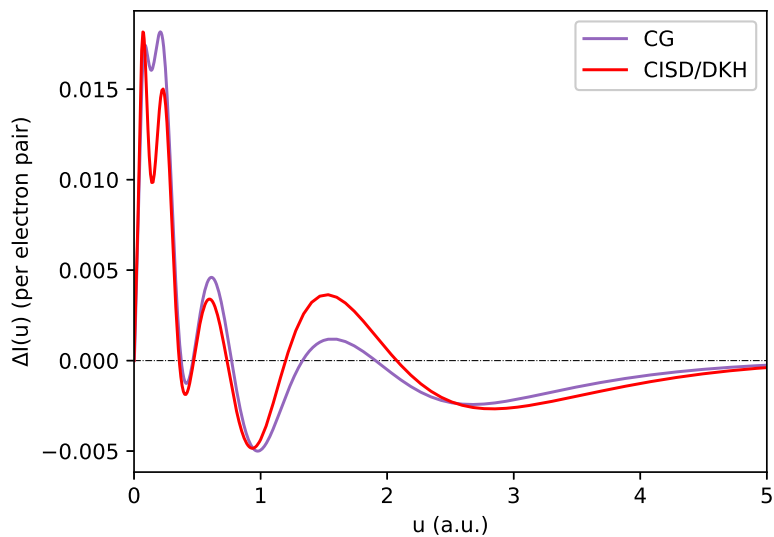}
\caption{Change on the rIPD due to scalar relativistic effects (using DKH approx.) for Xe atom computed at CISD level ($\Delta I(u) = I^{\textrm{rel}}(u)- I^{\textrm{nonrel}}(u)$). To facilitate the comparison, we have included as reference the change in $I(u)$ obtained by using the 4-component DHF calculation (including the CG interaction) and the non-relativistic HF one. In both cases, the results were obtained with Dyall-DZ basis~\cite{dyall2016relativistic}.}
\label{fig:xe_cg_dkhcisd}
\end{figure}

In order to have an approximate treatment of electronic correlation effects and to account for the scalar relativistic ones, we have chosen the DKH approximation for the Hamiltonian and the CISD approximation for $\Psi$ (see for example the results obtained for Xe atom in Fig.~\ref{fig:xe_cg_dkhcisd}). We have observed that at the correlated level there is a contraction of the inter-electronic distance, which is caused again by scalar-relativistic effects. Nevertheless, this result need to be confirmed by a full 4-component calculations including correlation effects.~\footnote{Let us remark that at HF level the inclusion of scalar relativistic effects with the DKH approximation produces a contraction in the mean value of the rIPD, but since we are interested in the role of electronic correlation effects we do not include this result.}

\begin{figure}[H]
\centering
\begin{subfigure}[t]{\textwidth}
\centering
\includegraphics[scale=0.8]{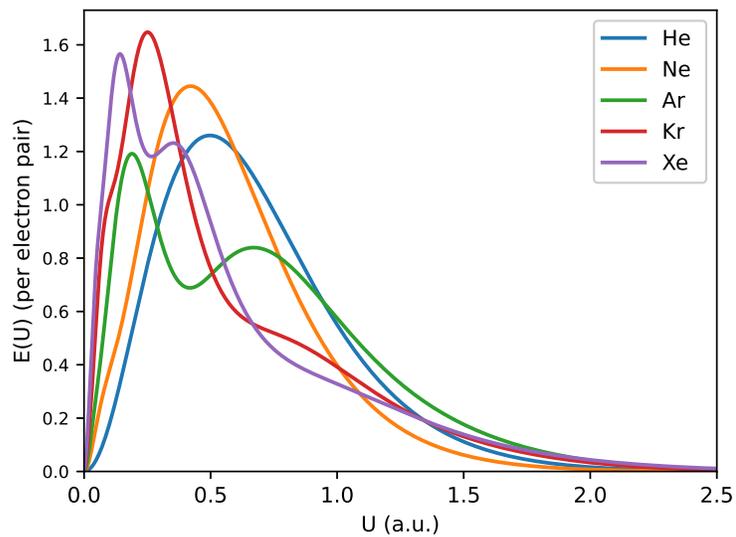}
\caption{}
\label{fig:all_cg_extraa}
\end{subfigure}
\hfill
\begin{subfigure}[t]{\textwidth}
\centering
\includegraphics[scale=0.8]{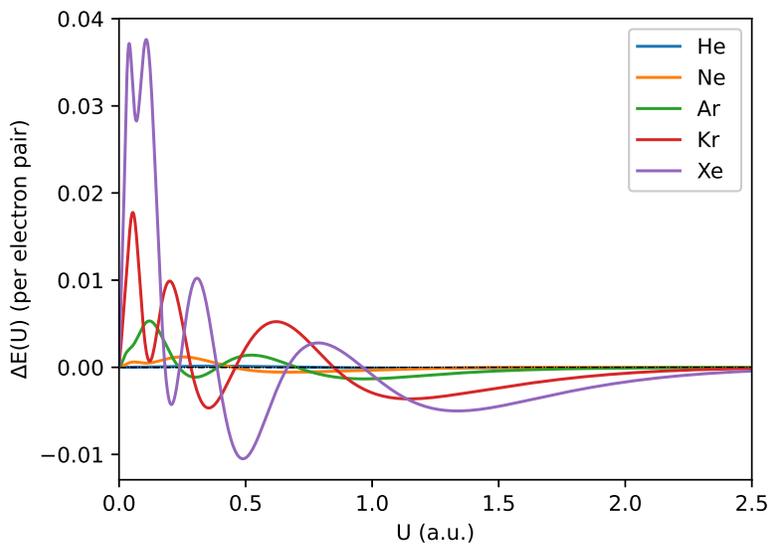}
\caption{}
\label{fig:all_cg_extrab}
\end{subfigure}

\caption{a) Relativistic rEPD for the noble gases (from He to Xe). b) Changes on the rEPD ($\Delta E(U) = E^{\textrm{rel}}(U)- E^{\textrm{nonrel}}(U)$) for the noble gases (from He to Xe) due to relativistic effects. Results obtained with Dyall-DZ basis~\cite{dyall2016relativistic}.}
\label{fig:all_cg_extra}
\end{figure}

\subsection{Relativistic effects on the rEPD.}
Next, we have studied the changes on the rEPD due to relativistic effects. The plot of the relativistic rEPD (see Fig.~\ref{fig:all_cg_extraa}) presents a pattern similar to the one obtained for the rIPD (plotted per electron-pair to fit the same scale). The main difference w.r.t.\ the rIPD plot is that the rEPD the probability is concentrated at lower values on the $U$-axis (it is almost located at half the distance of the rIPD) because the rEPD is related to the center of mass of the pair-probability distribution.
Concerning the change of the rEPD due to relativistic effects, we have also observed a contraction on it (see Fig.~\ref{fig:all_cg_extra}) that can also be quantify in terms of the mean value of the distribution (i.e.\ $\langle U \rangle = \int  dU U E(U)$), which is also tabulated in Table~\ref{tab:index}. Indeed, the analysis of the change on $\langle U \rangle $ due to relativistic effects indicates that a contraction is produced like in the rIPD, which can also be associated with a density driven effect. Moreover, the change on the variance of the rEPD (see $\sigma _U ^2$ in Table~\ref{tab:index}) reveals that the spread of the distribution is reduced when relativistic effects are considered (as it happened for the rIPD). Furthermore, we have also employed the DKH approximation and CISD calculations to confirm that the changes on the rEPD are also present when electronic-correlation effects are accounted.  
\begin{figure}[H]
\includegraphics[scale=0.8]{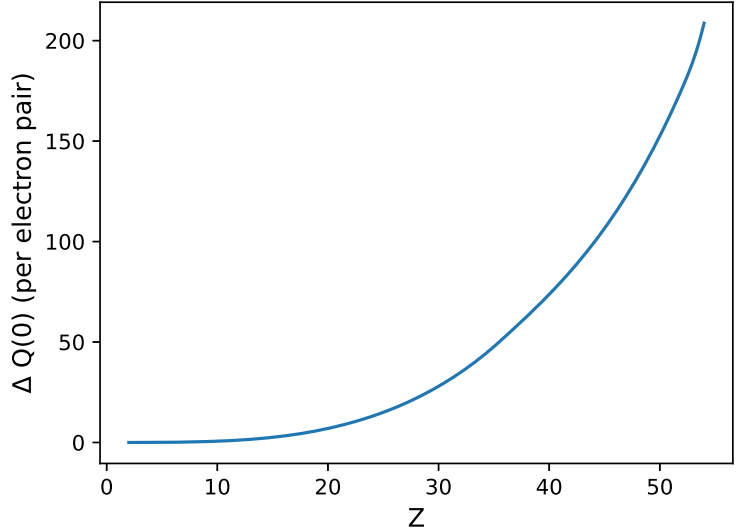}
\caption{Change on the electron-electron counter-balanced density, $Q({\bf 0})$, in atomic systems due to relativistic effects ($\Delta Q ({\bf 0}) = Q^{\textrm{rel}}({\bf 0})- Q^{\textrm{nonrel}}({\bf 0})$) against $Z$. Results obtained with Dyall-DZ basis~\cite{dyall2016relativistic}.}
\label{fig:all_cg_E0}
\end{figure} 
Finally, an interesting result is given by the analysis of the change in the electron-electron counter-balanced density due to relativistic effects in atomic systems. In Fig.~\ref{fig:all_cg_E0} we have plotted $\Delta Q ({\bf 0}) = Q^{\textrm{rel}}({\bf 0})- Q^{\textrm{nonrel}}({\bf 0})$ against the atomic number. Our results suggest an increase w.r.t.\ $Z$ of probability for the electrons to be at mirroring positions (${\bf r}_1 = -{\bf r}_2$) when forming electron pairs. That is to say, there is an increase in the PPD at $n_2({\bf r}_1,-{\bf r}_1)$ when relativistic effects are enhanced. {This can be explained at the mean-field level by an increase on the s$-$p and p$-$d exchange, which is (mainly) caused by two factors: a) the contraction of the orbitals that enhances two-electron interactions in general, and b) the contribution due to `opposite-spin' terms (see Eq.~\eqref{Eq:rel_PPD_2RDM}) when $Z$ gets larger. The s$-$p and p$-$d exchange, a.k.a.\ Fermi correlation, is of angular type and makes the electrons move to opposite sides of the nucleus~\cite{koga2002interelectronic,koga2002interelectronic2,hollett2012distributions,hollett2012distributions,todd2021measuring}}   

\section*{Conclusions}
Due to relativistic effects there is a contraction of radial density $D(r)$~\cite{pyykko1988relativistic}, which also reduces the mean inter-electronic distance $\langle u \rangle$ and displaces the center of mass of the PPD towards the origin. Indeed, the changes on the mean-value of the rIPD and of the rEPD can also serve to quantify the contraction of these distributions. Our results indicate that the scalar relativistic effects produce the majority of the changes on the rIPD and the rEPD, while the effect of using different\slash{}approximate electron-electron interactions is negligible. Therefore, we can conclude that the changes observed on the rIPD and rEPD are a density driven effect because the contraction of $D(r)$ towards the nuclei also reduces the inter-electronic distance and approaches to the origin the center of mass of the PPD. Moreover, the analysis of the decrease in $\langle u \rangle$ explains the increase of the Coulomb interaction $V^{\textrm{C}}_{ee}$.

Since the scalar relativistic effects dominate the changes on the rIPD and rEPD, we have performed post-HF calculations within the CISD approximation and including the DKH approximation (that provides an approximate treatment of the scalar relativistic effects) to show that we can expect similar changes on the rIPD and rEPD when electronic-correlation effects are considered. Nevertheless, post-DHF calculations using 4-component spinors (e.g.\ Kramers' restricted CISD, CCSD, among others.) should be performed to confirm this result. We are currently working along this line to obtain correlated-relativistic rIPDs. These relativistic correlated rIPDs should also be more suitable for comparison to experimental results than non-relativistic rIPDs (especially for heavy elements).  

In this work we have limited ourselves to the scalar part of the total 16$\times$16 rIPD (i.e.\ the trace of that matrix). The scalar part only allows for the calculation of the Coulomb part of the electron-electron interaction. With the 16$\times$16 rIPD also the effects on other components of the electron-electron interaction energy become accessible. Therefore, the generalization of the rIPD to a 16$\times$16 matrix could facilitate the study of the effects of relativity in more approximations to the electron-electron interaction; it could also serve to analyze the contribution in terms of the so-called large and the small components. This is an open question that will be studied in future works.

\section*{Acknowledgements}
M.R.-M. acknowledges the European Commission for an Horizon 2020 Marie Skłodowska-Curie Individual Fellowship (891647-ReReDMFT). K.J.H.G. acknowledges support by the Netherlands Organisation for Scientific Research (NWO) under Vici grant 724.017.001. M.R.-M. would also like to acknowledge Mr.~C. Chibueze, Dr.~S. Sitkiewicz,  Prof.~Dr.~P. Gori-Giorgi, Prof.~Dr.~J. Toulouse, and Prof.~Dr.~A. Savin for the fruitful discussions. 

\appendix

\begin{figure}[H]
\includegraphics[scale=0.8]{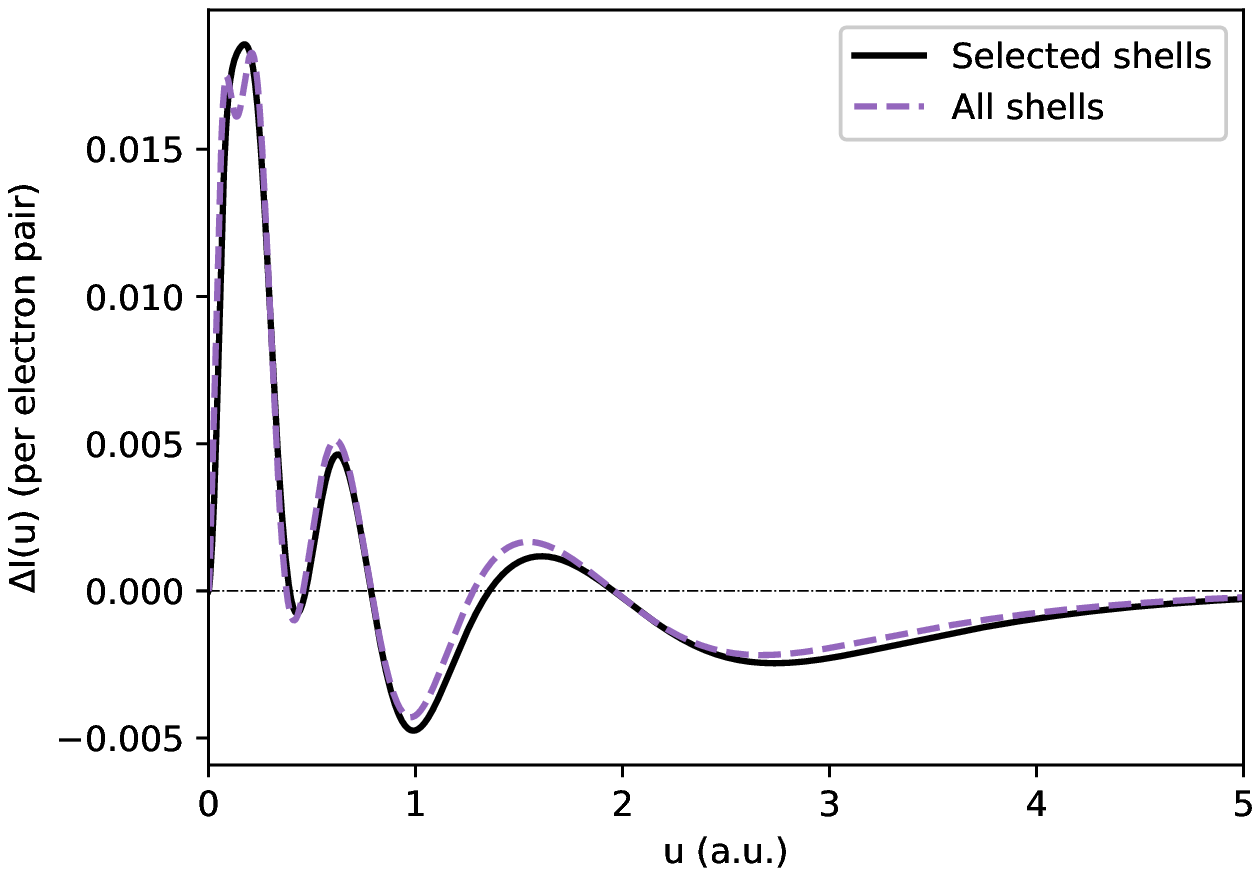}
\caption{Change on the rIPD on Xe due to relativistic effects for the following intra- and inter-shell contributions: 1s2s, 1s3p, 1s3d, 2s2p, 2s3s, 2p3p, 2p3d, 2p4s, 2p4p, 2p4d, 2p5s, 2p5p, 3s3d, 3p3d, 3p4s, 3p4p, 3p4d, 3d, 3d4s, 3d4p, 3d5s, 3d5p, 4s4d, 4s5s, 4s5p, 4p, 4p4d, 4p5s, 4p5p, 4d, 4d5s, 4d5p, 5s5p. We have included the total change on the rIPD for comparison. Results obtained with Dyall-DZ basis~\cite{dyall2016relativistic} and using the CG approximation for the electron-electron interaction.}
\label{fig:xe_shells_Dintra}

\includegraphics[scale=0.8]{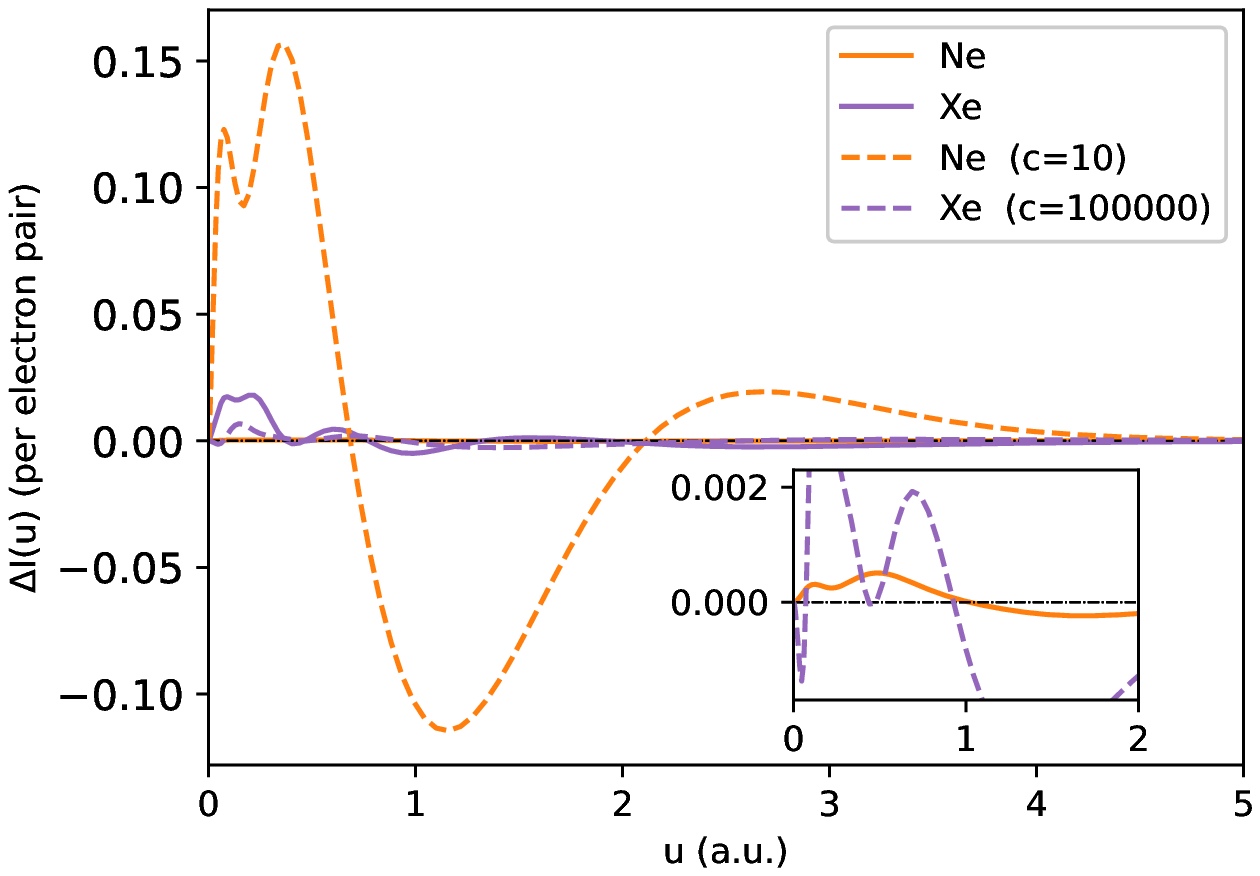}
\caption{Change on the rIPD due to relativistic effects (i.e.\ $\Delta I(u) = I^\textrm{rel} (u)- I^\textrm{nonrel} (u)$) with modified values of the speed of light ($c=10$\;a.u.\ for Ne atom and $c$=100\,000\;a.u.\ for Xe atom) plotted with dashed lines. We have included the result obtained without modifying the value of $c$ to facilitate the comparison. Results obtained with Dyall-DZ basis~\cite{dyall2016relativistic}.}
\label{fig:ne_xe_c}
\end{figure}

\section{Inter- and intra-shell contributions to the rIPD.}
\label{appexA}
The changes on the rIDP account for several inter- and intra-shell contributions~\footnote{For this work, the shells were defined according to the non-relativistic shell structure.}. We have studied separately the inter- and intra-shell changes due to relativistic effects on the Xe atom. Let us start by noticing that the Xe atom contains 54 electrons and 11 shells (i.e.\ 1s, 2s, 2p, 3s, 3p, 3d, 4s, 4p, 4d, 5s, and 5p). The intra-shell contributions to the total rIPD are given by the electron pairs formed within each shell. On the other hand, the inter-shell contributions are given by the electron pairs formed among different shells (e.g.\ pairs formed between the 1s and 2s electrons). We have labeled the intra-shell contributions using the names of the shells (i.e.\ 1s, 2s, \ldots, etc.) while for the inter-shell contributions we have used the labels 1s2s, 1s2p, \ldots. In the end, 66 (inter and intra) shell contributions produce the global change of the rIPD. Interestingly, with only 33 out of the 66 terms an accurate shape of the relativistic effect on the rIPD can be produced (see Fig.~\ref{fig:xe_shells_Dintra}a). This result implies that not all shells contribute equally; but it also shows that there is not a major contribution from a particular shell. However, it is shown that the major changes on the rIPD are of inter-shell type. We have also noticed that the first peak observed on $\Delta I(u)$ at very short inter-electronic distances is mostly due to changes involving the 1s shell (intra- and inter-shell contributions), where the electrons are in average at shorter inter-electronic distances than in the rest of shells.

We have confirmed that the shell contributions to the Coulomb electron-electron interaction ($\Delta V_{ee} ^\textrm{C}=V_{ee} ^\textrm{C,rel}-V_{ee} ^\textrm{C,nonrel}$) is mainly given by inner shells, which is due to the increase in the proximity of the electrons near the nuclei w.r.t.\ to the electrons located at the valence shell.

\begin{figure}[H]
\centering
\includegraphics[scale=0.8]{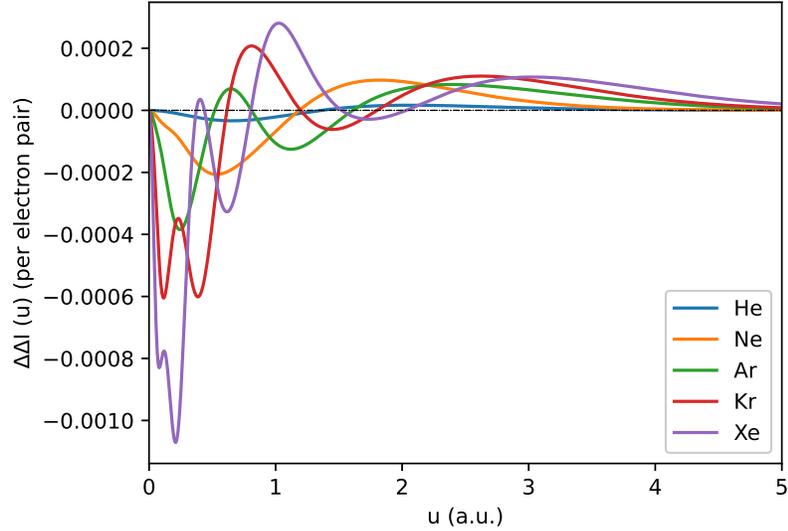}
\caption{Change of the rIPD when a partial neglect of the small-component electron-repulsion integrals (i.e.\ LVNEW option in DIRAC~\cite{visscher1997approximate,saue2020dirac} program) is used w.r.t.\ the full CG interaction (i.e.\ $\Delta \Delta I(u)  = I^{\textrm{LVNEW}}(u) - I^{\textrm{CG}}(u)$) for the noble gases. Results obtained with Dyall-DZ basis~\cite{dyall2016relativistic}.}
\label{fig:he_xe_lv_cg}
\end{figure}

\section{Increase on the relativistic on-top pair density.}
\label{appexB}
{
\newpage
\vspace*{-3cm}\begin{table}[H]
\resizebox{6.5cm}{11cm}{
    \caption{Integrated Hartree--Fock (HF) and Dirac--Hartree--Fock (DHF) on-top pair densities (i.e.\ $P^{\textrm{nonrel}}({\bf 0})$ and $P^{\textrm{rel}}({\bf 0})$, respectively) computed using the Dyall-DZ basis~\cite{dyall2016relativistic}}.
    \label{tab:index2}
    \begin{tabular}{crr}\hline
    {System} & $P^{\textrm{nonrel}}({\bf 0})$ & $P^{\textrm{rel}}({\bf 0})$ \\ \hline
    \multicolumn{3}{c}{Atoms} \\
 He	&   0.190	& 0.190\\
 Li	&   0.784	& 0.785\\
 Be	&   2.099	& 2.100\\
 F	&  29.968	&30.130\\
 Ne	&  42.536	&42.820\\
Na	&  58.490	&58.968\\
Mg	&  78.294	&79.060\\
Cl	& 249.362     &  254.530 \\
Ar	& 301.377     &  308.425 \\
K	& 360.594     &  370.073 \\
Ca	& 427.438     &  439.987 \\
Zn	&1599.304     & 1711.725 \\
Br	&2648.435     & 2908.834 \\
Kr	&2904.603     & 3208.423 \\
Rb	&3465.391     & 3531.078 \\
Sr	&3724.362     & 3877.555 \\
Cd	&7749.992     & 8925.244 \\
I  &   10456.535 &    12888.590 \\
Xe  &   11078.534 &    13830.304 \\
    \multicolumn{3}{c}{hydrides} \\
H$_2$ &      0.046 &     0.046 \\
LiH  &       0.784 &	 0.802 \\
NaH  &      58.490 &       58.982 \\
KH  &     360.594 &      370.076 \\
RbH  &    3465.391 &     3531.051\\ \hline
    \end{tabular}
}
\end{table}
In Eq.~\eqref{Eq:IPD} $P({\bf 0})$ corresponds to the integrated on-top pair density (with $n_2({\bf r},{\bf r})$ being the on-top pair density), which according to Table~\ref{tab:index2} increases at the DHF level w.r.t.\ the HF one. 
}  

\section{Change in the rIPD due to relativistic effects when the speed of light is modified.}
\label{appexC}  
In Fig.~\ref{fig:ne_xe_c} we have plotted the change in the rIPD due to relativistic effects when the value of the speed of light is modified. For Neon atom the changes on the rIPD reveal an important increase of probability at shorter inter-electronic distances. On the contrary, for Xe atom a clear reduction of the relativistic effects in the rIPD is obtained as it approaches the non-relativistic result.

\section{More approximations to the electron-electron interactions.}
\label{appexD}
The scalar effects dominate the changes observed on the rIPD when relativistic effects are considered. However, in this work we have also analyzed the effect of using other approximations to the electron-electron interaction. Different combinations have been studied and in all cases very small changes w.r.t.\ our best approximation (i.e.\ the one based on the CG interaction) were obtained. Indeed, even when ``strong'' approximations are employed, the changes observed on the rIPD due to relativistic effects are very similar to the results obtained with the CG interaction. For example, the use of the Coulomb interaction with neglect of integrals that only involve small-component functions (i.e.\ using the LVNEW option in DIRAC program) leads to very small differences in rIPD w.r.t.\ the one obtained with the full CG approximation (see Fig.~\ref{fig:he_xe_lv_cg}). We have plotted this difference in Fig.~\ref{fig:he_xe_lv_cg}, where we can see that their effect is lower than the effect due to the scalar-relativistic effects and they only introduce small changes in the rIPD. Let us remark that in all cases the DHF equations were solved self-consistently; thus, the self-consistent spinors differ when different electron-electron interaction approximations are employed. 
 
For this particular example, we observe from Fig.~\ref{fig:he_xe_lv_cg} that the using the LVNEW option reduces the amount of contraction of the inter-electronic distance w.r.t.\ the CG interaction. The reason for this difference goes beyond the scope of the present work and may require an analysis using the 16$\times$16 component 2-RDM to study also the corresponding currents that enter the Gaunt interaction expression.

\bibliography{general}       


\end{document}